\documentclass[aps,prl,twocolumn,a4paper,10pt,notitlepage,footinbib,superscriptaddress,longbibliography]{revtex4-1}
\usepackage[english]{babel}
\usepackage{lmodern}
\usepackage[utf8]{inputenc}
\usepackage{endnotes}
\usepackage{amssymb,amsmath,amsfonts}
\usepackage{textcomp}
\usepackage{braket}
\usepackage{ulem}
\usepackage{soul}
\usepackage{graphicx}
\usepackage{dcolumn}
\usepackage{bm}
\usepackage{xcolor}
\usepackage{soul}
\usepackage{xr}
\usepackage{cleveref}

\definecolor{crimson}{rgb}{0.75686,0,0.262745}
\definecolor{saphire}{rgb}{0.0,0.196,0.372549}
\definecolor{plum}{rgb}{0.50588,0.007843,0.3843137}

\newcommand{\qtens}{{\bf{Q}}}
\newcommand{\identitytens}{{\bf I}}

\begin{document}

\title{3D active nematic disclinations behave as Majorana quasiparticles}

\author{Louise C.~Head$^{*}$}
\affiliation{SUPA, School of Physics and Astronomy, University of Edinburgh, Peter Guthrie Tait Road, Edinburgh, EH9 3FD, UK}
\author{Giuseppe Negro$^{*}$}
\affiliation{Dipartimento di Fisica, Università degli Studi di Bari and INFN, Sezione di Bari, via Amendola 173, Bari, I-70126, Italy}
\author{Livio N.~Carenza}
\affiliation{Faculty CS Physics, Koc University, Rumelifeneri Yolu 34450 Sariyer, Istanbul,  Turkey}
\author{Ryan R. Keogh}
\affiliation{SUPA, School of Physics and Astronomy, University of Edinburgh, Peter Guthrie Tait Road, Edinburgh, EH9 3FD, UK}
\author{Giuseppe Gonnella}
\affiliation{Dipartimento di Fisica, Università degli Studi di Bari and INFN, Sezione di Bari, via Amendola 173, Bari, I-70126, Italy}
\author{Alexander Morozov}
\affiliation{SUPA, School of Physics and Astronomy, University of Edinburgh, Peter Guthrie Tait Road, Edinburgh, EH9 3FD, UK}
\author{Enzo Orlandini}
\affiliation{Department of Physics and Astronomy, University of Padova and  INFN, Sezione Padova, Via Marzolo 8, I-35131 Padova, Italy}
\author{Tyler N.~Shendruk}
\affiliation{SUPA, School of Physics and Astronomy, University of Edinburgh, Peter Guthrie Tait Road, Edinburgh, EH9 3FD, UK}
\author{Adriano Tiribocchi}
\affiliation{Istituto per le Applicazioni del Calcolo, Consiglio Nazionale delle Ricerche, via dei Taurini 19, Roma, 00185, Italy}
\author{Davide Marenduzzo}
\affiliation{SUPA, School of Physics and Astronomy, University of Edinburgh, Peter Guthrie Tait Road, Edinburgh, EH9 3FD, UK\\
$^*$Equal contribution}

\begin{abstract}
Quasiparticles are low-energy excitations with important roles in condensed matter physics. An intriguing example is provided by Majorana fermions, quasiparticles which are identical to their antiparticles. Despite being implicated in neutrino oscillations and topological superconductivity, their experimental realisations remain scarce. Here we propose a purely classical realisation of Majorana fermions, in terms of $3$-dimensional disclination lines in active nematics. Activity is required to overcome the elastic cost associated with these excitations,  so they can appear in steady state. We combine topology and simulations to show that active nematics under confinement spontaneously create in their interior topologically charged disclination lines and loops, akin to Majorana quasiparticles with finite momentum. Within an elongated channel, we find a phenomenology similar to that of the Kitaev chain, as local Majorana-like excitations appear near surfaces, while a non-local system-spanning helical disclination line can arise along the centre. In unconfined active turbulence, Majorana-like charged loops are instead exceedingly rare, suggesting that boundaries are crucial to generate these quasiparticles, as in quantum condensed matter. We suggest that 3-dimensional active disclinations can be used to probe the physics of Majorana spinors at a much larger scale than traditionally considered, potentially facilitating the experimental observation of their dynamics.
\end{abstract}


\maketitle



Liquid crystals have long served as a fruitful playground where abstract ideas from mathematics, condensed matter, cosmology and particle physics can find a practical application~\cite{mermin1979,mermin1989,chuang1991}. A well-known example is the topological theory of defects and disclinations in nematics and cholesterics in $2$ and $3$ dimensions, which provides a tangible application of homotopy theory~\cite{mermin1989,lavrentovich2001}. Another instance is given by topological solitons, such as skyrmions, torons and hopfions, which behave as quasiparticles with rich and fascinating emerging phase behaviour and dynamics~\cite{ackerman2017,poy2022,wu2022,carenza2022,shankar2022,bowick2022}. With respect to other condensed matter systems such as superconductors and superfluids, which also harbour non-trivial topological phases and quasiparticles~\cite{kitaev2001,qi2011,alicea2012,wolfle2018}, the spatiotemporal scales of liquid crystal patterns are usually much larger -- a fact that can be useful to simplify their experimental studies. 

\begin{figure}[h!]
\includegraphics[width=\columnwidth]{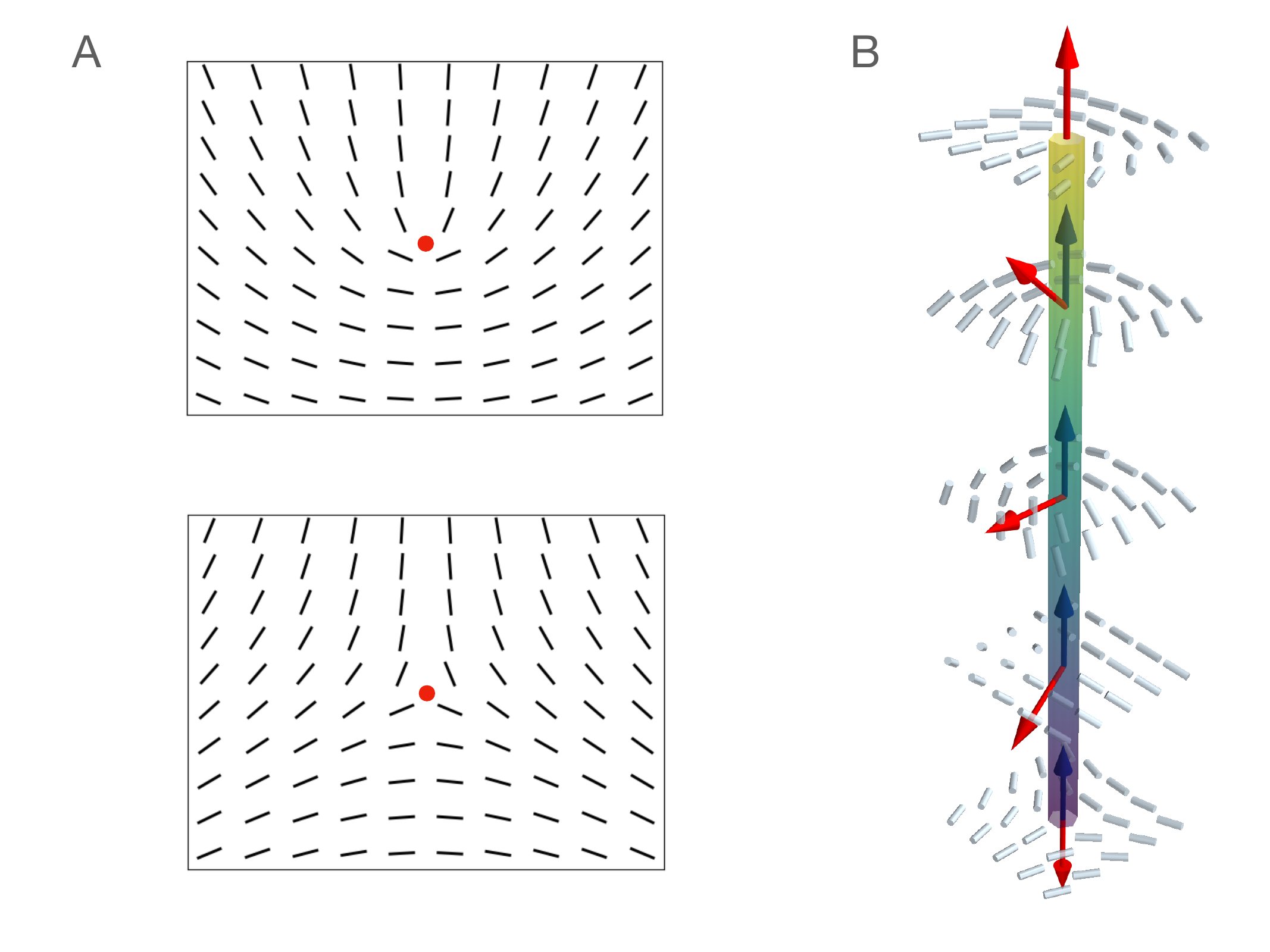}
\caption{A) Local planar director profile for topological charge $s=1/2$ (top) and $s=-1/2$ (bottom) disclinations, with the defect shown as a red dot. The two patterns are distinct in 2D, but they can be mapped onto each other when embedded in 3D. Either profile can be viewed as a Majorana quasiparticle at rest. (B) A disclination line containing a $+1/2$ and a $-1/2$ local defect profile, which is therefore equivalent to the trivial (defect-free) nematic state. This line can therefore be viewed as a pair of Majorana quasiparticles. The colouring indicates the value of $\cos \beta={\mathbf T}\cdot {\mathbf \Omega}$, with ${\mathbf T}$ (corresponding to the blue arrows) and ${\mathbf \Omega}$ (red arrows) respectively the local tangent and rotation vector~\cite{schimming2022,adriano2024} for each point on the disclination.}
\label{fig1}
\end{figure}

Here we show that liquid crystals possess an intriguing qualitative analogue of Majorana excitations, which are 
quasiparticles, that are equivalent to their corresponding antiparticles~\cite{majorana1937}. Specifically, we show that the liquid crystal analogue of a Majorana particle is a 3-dimensional and topologically charged nematic disclination loop. The qualitative analogy holds because a $+1/2$ topological defect profile can be smoothly transformed, in 3 dimensions, into a $-1/2$ profile (see e.g.~\cite{copar2013}, and Fig.~\ref{fig1}). More formally, we show that this profile can be mapped to a spinor solution of the Majorana equation in $1+1$ dimensions. Several condensed matter examples of Majorana quasiparticles have been found in the last two decades, for instance in the Kitaev chain made up of spinless electrons~\cite{kitaev2001,alicea2012}, or in systems of magnetic skyrmions~\cite{mohanta2021}. However, previous realisations were always quantum, whereas the nematic analogue we propose is classical, which is appealing as it provides the tantalising opportunity to observe quantum mechanical-like behaviour at the mesoscale. Whilst in passive nematic liquid crystals such Majorana-like excitations cost elastic energy, so that the associated spectrum is gapped, in active nematics~\cite{ramaswamy2010,marchetti2013,doostmohammadi2018,duclos2020,binysh2020} they arise spontaneously due to the continuous energy input into the system, 
so that they may be viewed as gapless (or almost gapless) excitations.

The physics of Majorana excitations~\cite{kauffman2022} suggests that, if they can be stably produced in the lab, these quasiparticles can be braided and entangled in a controllable way, which may be useful for topological computing~\cite{kitaev2003,lian2018}. Indeed, liquid crystals have been recently proposed as an alternative medium to create new-generation computers and information processors~\cite{kos2022}; the mapping we propose in this work may provide further motivation to explore the potential of liquid crystals as topological computers. The Majorana mapping we discuss may also shed light, in turn, on the physics of active turbulence~\cite{alert2022,giomi2015} in 3D, which has recently gained attention and has been experimentally realised in~\cite{duclos2020}. Specifically, we suggest that the onset of turbulence is linked to a transition between a trivial and a topological phase, where non-local topological excitations spontaneously appear. Topological excitations and defects would then potentially play an important role in the physics of turbulence in active apolar nematics -- at odds with the case of polar nematics in 2D, where the underlying scaling laws can be obtained without considering defects~\cite{alert2020}.\\
 
\section*{Mapping between Majorana particles and nematic disclination lines}

To begin with, we review the physics of Majorana spinors in $1+1$D, following~\cite{kauffman2022}. 
To construct them, it is customary to start from a Clifford algebra, which in $1+1$D can be built from the following two matrices (or generators),
\begin{equation}
\epsilon = \begin{pmatrix}
-1 & 0 \\
0 & 1 
\end{pmatrix}, \qquad 
\eta = \begin{pmatrix}
0 & 1 \\
1 & 0
\end{pmatrix}.
\end{equation} 

The algebra is such that the generators anticommute and have square equal to unity~\cite{kauffman2022,pal2011},
\begin{equation}
\epsilon^2=1, \qquad \eta^2=1, \qquad \epsilon\eta+\eta\epsilon=0.
\end{equation}
Note that the associated Clifford algebra is Cl(2,0), 
as both $\epsilon$ and $\eta$ have a square which is positive (and equals unity). 

Let us define $\psi_0 = e^{i(px-Et)}$, a function associated with a free particle of mass $m$, momentum $p$, and energy $E$. If we write the Majorana operator as~\cite{kauffman2022}
\begin{equation}\label{Majoranaeq}
D = i \eta \frac{\partial}{\partial t} + i \eta \epsilon \frac{\partial}{\partial x} - i \epsilon m,
\end{equation}
we find 
\begin{equation}
D \psi_0 = 
\left[\eta(E-\epsilon p)-i\epsilon m \right] \psi_0
\equiv (A+iB) \psi_0.
\end{equation}

It can be seen that $A+iB$ is nilpotent, as $(A+iB)^2=E^2-p^2-m^2=0$. As a result, the field $\psi = (A+iB)\psi_0$ provides a set of solutions for the Majorana equations in $1+1$D. The two spinor solutions are the two-column vectors in the following matrix,
\begin{equation}\label{majorana1d}
\psi = \begin{pmatrix}
-m \sin(\theta) & (E-p)\cos(\theta) \\
(E+p)\cos(\theta) & m\sin(\theta) 
\end{pmatrix},
\end{equation}
with $\theta=px-Et$.
For a particle at rest ($p=0$), these solutions can be written as 
\begin{equation}\label{majorana1drest}
\psi = m \begin{pmatrix}
-\sin(\theta) & \cos(\theta) \\
\cos(\theta) &  \sin(\theta)
\end{pmatrix},
\end{equation}
where now $\theta=-mt$. 

Let us now consider a disclination line in a nematic liquid crystal. We consider a toroidal surface encircling the disclination line, which is topologically equivalent to $S_1\times S_1$. Following~\cite{copar2013}, we call $u$ the angle parametrising the position along the smaller meridian $S_1$ circle, perpendicular to the disclination line, and $v$ the angle determining the position along the larger longitudinal $S_1$ circle, tangential to the disclination line. At $v=0$, the defect pattern can be described by a single angle $\alpha(u)$, so that
\begin{equation}\label{nematicu}
{\mathbf n}=(\cos(\alpha(u)),\sin(\alpha(u))),
\end{equation}
where the two components are along ${\mathbf m}$ and ${\mathbf T}\times{\mathbf m}$ respectively, where ${\mathbf T}$ is the unit vector in the tangential direction to the disclination line, and ${\mathbf m}$ is the unit vector in the perpendicular plane which corresponds to the direction of the nematic director field at $u=0$. 

The local director field pattern in Eq.~(\ref{nematicu}) with $\alpha(u)=-1/2 u$, corresponding to a triradius, can be seen as a Majorana-like quasiparticle at rest, as the solution is formally the same as that in Eq.~(\ref{majorana1drest}) (with $\theta=\alpha(u))$. 
Physically, the identification of a triradius as a Majorana-like quasiparticle is motivated by the observation that its antiparticle, the comet-like defect with $\alpha(u)=+1/2 u$, is equivalent to it because one can be transformed into the other by rotating the director by $\pi$ around an angle in the plane perpendicular to ${\mathbf T}$ (Fig.~1B, and~\cite{copar2013}).

Whilst the local profile in Eq.~(\ref{nematicu}) can be viewed as a Majorana-like quasiparticle {\it at rest}, a closed disclination loop may be viewed as its counterpart with $p\ne 0$, qualitatively equivalent to the spinor solutions in Eq.~(\ref{majorana1d}). It is well known that disclination loops in nematics can carry a topological charge $Q$, which is conserved modulo $2$~\cite{mermin1979,lavrentovich2001}. For loops which are not pierced by other disclinations, there are only two possible topological classes. The first class is charged disclinations, which are topologically non-trivial, have $Q=1$, and are equivalent to a $+1/2$ or a $-1/2$ profile transported over the loop. These are qualitatively equivalent to a Majorana particle in $1+1$D with finite momentum, or equivalently to a spinor solution of Eq.~(\ref{majorana1d}). The second class is uncharged loops with $Q=0$, which are topologically trivial, and contain, for instance, one $+1/2$ profile and a $-1/2$ profile. Uncharged loops can be shrunk to leave a defect-free state. These loops are equivalent to a pair (or an even number) of Majorana quasiparticles, which annihilate with each other. 

\if{While the mapping we just described involves the representation of Majorana quasiparticles as $2\times 1$ spinors, we may also choose a representation in terms of matrices. Here, Majorana elements may be identified with the generators of a Clifford algebra. In the case of the particle at rest, the Clifford algebra is $Cl(1,0)$, and we can choose to represent, for instance, the (Majorana) planar defect configuration in Fig.~\ref{fig1}a via the Pauli matrix $\sigma_x=\eta$, such that the Clifford algebra is composed by $1$ and $\sigma_x$. When the local profile is transported along a loop, corresponding to a Majorana quasiparticle with $p\ne 0$, the relevant matrices are $\sigma_x$, which we can view as the Majorana particle, and $\sigma_z=-\epsilon$, which is a parity transformation mapping the particle onto its antiparticle. This representation is isomorphic to $Cl(2,0)$, which is the representation we used to write the Majorana operator in Eq.~(\ref{Majoranaeq}).}\fi

Disclination loops -- whether charged, hence Majorana-like, or not -- cost elastic energy, scaling as $K l$, with $K$ an elastic constant and $l$ the disclination size. Given typical values of $K$ in liquid crystals are $\sim 10$ pN, a $l\sim 1$ $\mu$m loop costs $\sim 1.8 \times 10^3 k_BT$, and is therefore a gapped excitation. Disclinations are therefore not seen in practice in a nematic liquid crystal when this is close to its ground state. One promising way to obviate this issue is to use active liquid crystals~\cite{ramaswamy2010,marchetti2013,doostmohammadi2018} instead of passive ones. In active nematics, such as microtubule-kinesin or actomyosin mixtures, ATP-driven motors (kinesin or myosin) drive the system out of equilibrium and provide a source of non-thermal fluctuations with energy much larger than $k_BT$, leading to spontaneous appearance of defects and disclination lines or loops in steady state~\cite{marchetti2013}. In other words, disclinations are gapless (or almost gapless) in active liquid crystals, and activity, which measures the density and strength of dipolar forces exerted by the molecular motors, can be thought of as an effective temperature, or inverse chemical potential. In what follows, we will therefore analyse various active nematic systems, and study the appearance and topology of local defect profiles and disclination lines and loops in each of these, commenting on the appearance of Majorana-like features.

\section*{Results and Discussion}

\subsection*{Confined active nematics: Majorana-like loops in double emulsions}
\begin{figure}[h!]
\includegraphics[width=\columnwidth]{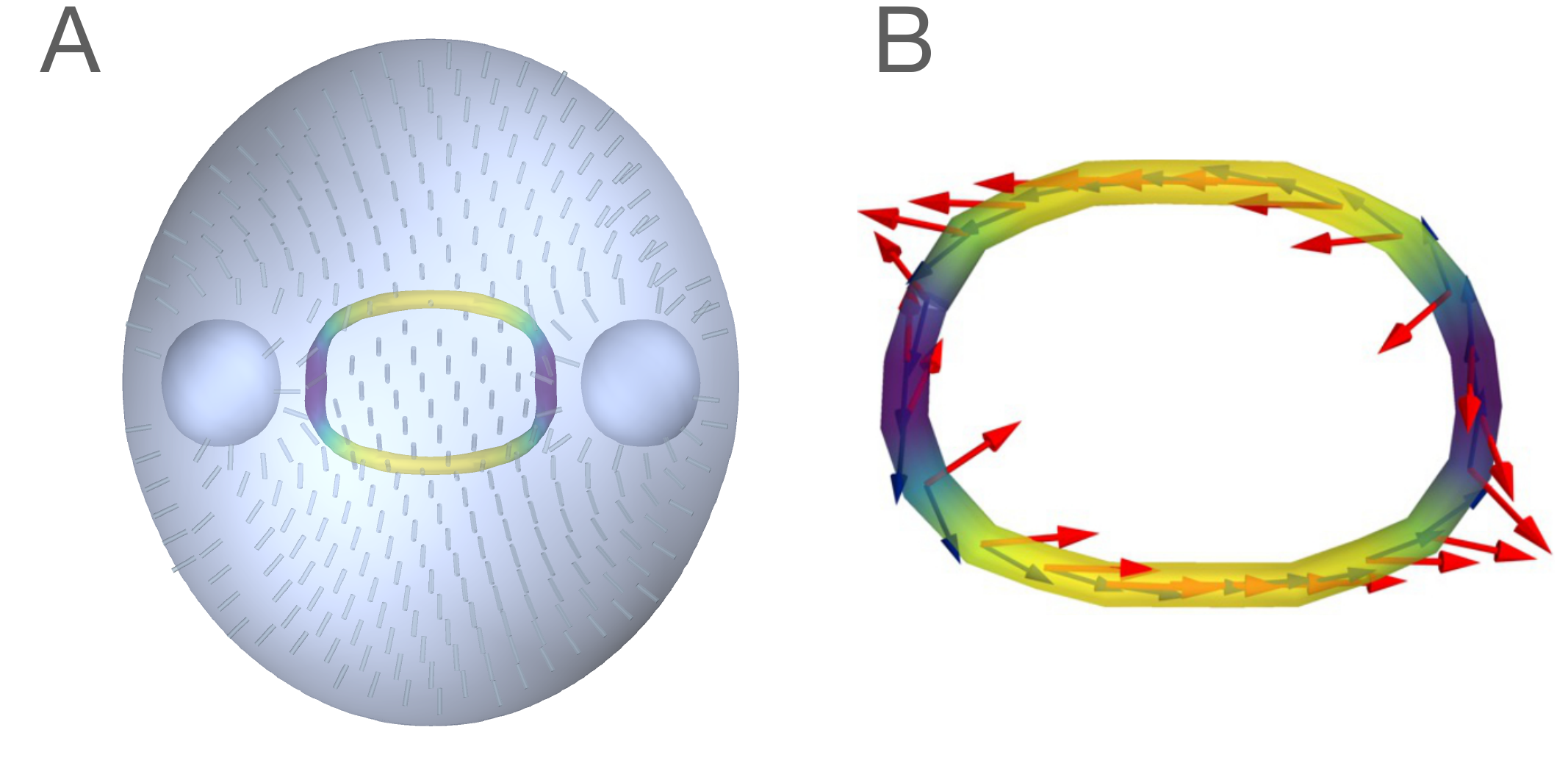}
\caption{Director field pattern and disclination lines (coloured according to the local $\cos{\beta}$ value, as in Fig.~\ref{fig1}), for an active nematic double emulsion, made up by inserting two isotropic passive droplets inside an active nematic droplet~\cite{adriano2024}. A topologically charged (+-+-) loop can be seen, corresponding to two changes back and forth from $+1/2$ to $-1/2$ local patterns, and visible via the colour corresponding to the local value of $\cos{\beta}$.} 
\label{fig2}
\end{figure}

We start by considering an active double emulsion (see Methods and~\cite{adriano2024}). This system is constituted by an active extensile nematic liquid crystal droplet which contains two smaller isotropic (passive and non-coalescing) droplets in its interior (Fig.~\ref{fig2}A). All droplets are deformable and we consider normal anchoring on all surfaces. Because the internal droplets are equivalent to two topological point charges (`hedgehogs'), the overall geometry requires the director field pattern to be topologically non-trivial globally -- i.e., it needs to include at least one charged, Majorana-like, disclination loop. 

Fig.~\ref{fig2} shows an example of the most likely topological pattern observed at intermediate activity, as the system starts to flow spontaneously due to activity~\cite{adriano2024}. This pattern consists of a single rotating Majorana-like charged loop which bridges the two passive droplets. This Majorana-like loop, which as previously mentioned is topologically protected due to the emulsion geometry, features two parity transformations~\cite{copar2013} back and forth from local $+1/2$ comet to $-1/2$ triradius profiles (Fig.~\ref{fig2}B). Hence, we call this a +-+- loop; a Saturn ring embracing a colloid in a passive liquid crystal~\cite{stark2001} would therefore be a - loop in this nomenclature. A single back-and-forth change (a +- loop) is instead equivalent to a topologically trivial loop~\cite{copar2013}. In terms of topological fermions, a +-+- loop can be viewed as a combination of three Majorana quasiparticles, which is topologically non-trivial. Each Majorana quasiparticle can then be more precisely identified with a $2\pi$ rotation of the ${\mathbf \Omega}$ vector (red arrows in Fig.~\ref{fig2}B). One of the rotations is the overall rotation of ${\mathbf \Omega}$ around the loop contour, the other two correspond to the two parity transformations.

While a charged loop is topologically required by the double emulsion geometry, activity extends the loop size such that it bridges the two passive droplets. Notably, the passive droplet surfaces attract $-1/2$ local profiles, in analogy with colloidal particles which are surrounded by Saturn rings, which are everywhere equivalent to a $-1/2$ profile. Unlike the case of colloids in a passive liquid crystal though, the loop in this case is dynamic (and rotates in the case shown in Fig.~\ref{fig2}). 

\subsection*{Active nematic pipes and an analogue of the Kitaev chain}

To further explore the validity of the mapping between defect profiles and Majorana quasiparticles, we now consider a quasi-1D system where the nematic is confined along two directions, rather than inside a spherical droplet. Specifically, we consider a nematic inside a long square parallelepiped, with degenerate planar anchoring on the boundaries, and periodic boundary conditions along the long direction~\cite{keogh2022} (see Methods for the underlying free energy and computational details). The motivation is to seek a liquid crystalline analogue of the Kitaev chain, which appears in a 1D geometry when Majorana particles couple to yield spinless fermions~\cite{kitaev2001}. In the Kitaev chain, there is a topological transition between a trivial strong-coupling state, where Majorana particles pair up locally in a way which is reminiscent of Cooper pairs in superconductors, and a non-trivial fermionic state made up of Majorana operators localised at the chain edges, which delocalise close to the transition~\cite{alicea2012}.

\begin{figure*}[t!]
\centering
\includegraphics[width=\textwidth]{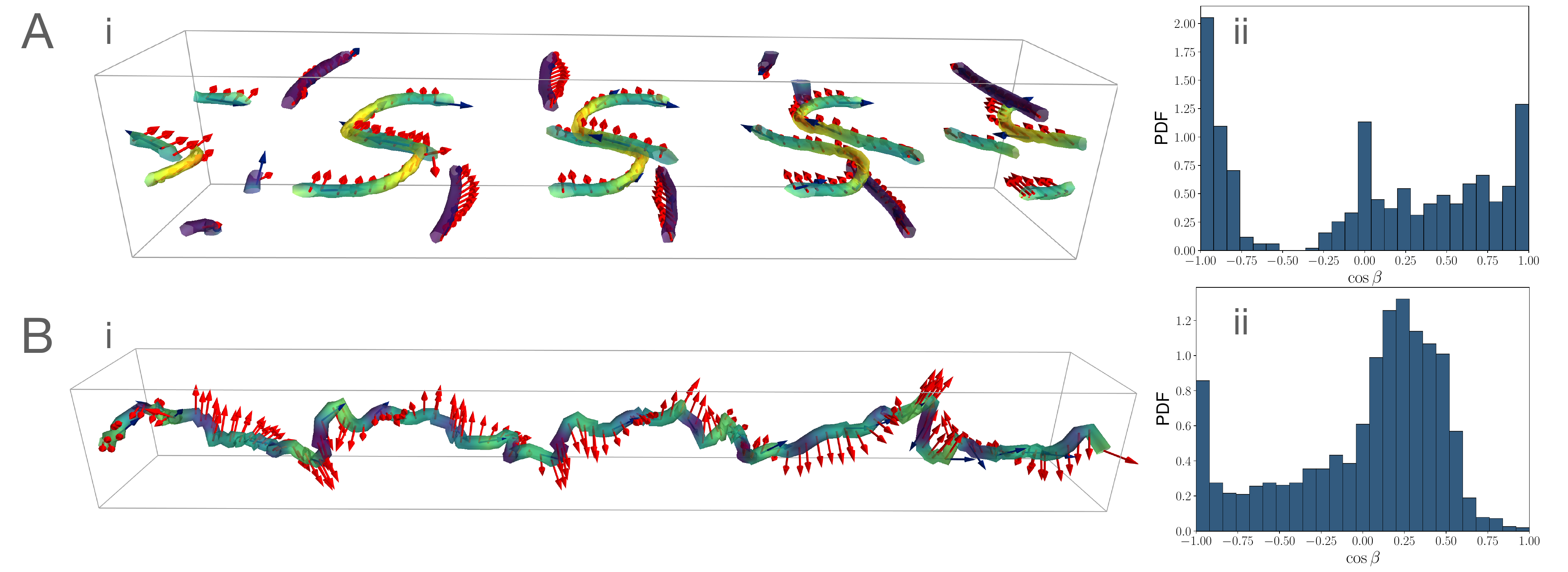}
\caption{(A) (i) Director field pattern and disclination lines (coloured according to the local $\cos{\beta}$ value, as in Fig.~\ref{fig1}), for the vortex lattice first described in~\cite{keogh2022}. (ii) Corresponding histogram of local values of $\cos{\beta}$, showing the abundance of $\pm 1/2$ Majorana-like local patterns. (B) (i) Director field pattern and disclination lines (coloured according to the local $\cos{\beta}$ value) for the double helix state of\cite{keogh2022}. (ii) Corresponding histogram of local values of $\cos{\beta}$.} \label{fig3}
\end{figure*}

In our active nematic pipe, for low activity, the ground state is 
defect-free, which, unlike for the active emulsion case, is compatible with the boundary conditions. 
Here, Majorana loops are gapped excitations as in a passive liquid crystal, and cannot be seen. For sufficiently large activity, a ``vortex lattice''~\cite{keogh2022} spontaneously appears, with defects in steady state (Fig.~\ref{fig3}A), accompanied by active flow. At early times, the emerging defects appear as uncharged disclination loops (Fig.~S1). Later on, charged disclination arcs of finite size stick to the boundary, corresponding to Majorana quasiparticles. The local value of $\cos{\beta}$ calculated from the disclination density tensor [see Materials and Methods, Eq.~(\ref{Dij})], peaks around $-1$ and $+1$, corresponding to local triradius and comet profiles respectively. The former localise at corners, and the latter most commonly stick to opposite walls and are stretched by the flow (Fig.~\ref{fig3}A). This localisation of defects close to surfaces is qualitatively reminiscent of Majorana excitations found deep in the topological phase of the Kitaev chain, where they have finite size and localise at the chain boundaries~\cite{alicea2012}. 


A qualitatively different structure is associated with the ``double helix'' state~\cite{keogh2022}, which appears for parameters for which the liquid crystal is in the isotropic phase in the passive limit (see Methods). Here, defects cost less free energy, and for sufficiently large activity  
the system self-organises to create a helical flow state with the main velocity component along the direction of the channel and with a defined handedness (here left-handed) arising through spontaneous chiral symmetry breaking. This structure is associated with an extended chiral disclination line with the same handedness, which can be viewed as a gapless excitation (Fig.~\ref{fig3}B). This is equivalent to a delocalised topological state in the Kitaev chain, which in the latter system is observed just at the transition between the trivial and topological phases~\cite{alicea2012}. The global topological charge of the extended disclination is zero, as the pattern is characterised by an even winding of the rotation vector ${\mathbf \Omega}$ around the tangent to the disclination line. Note that, because its global topological charge is zero, this disclination line is equivalent to an even number of Majorana quasiparticles. This is again qualitatively similar to the situation in the Kitaev chain, where with periodic boundary conditions Majorana fermions do not unpair~\cite{kitaev2001,alicea2012}.

\subsection*{Unconfined active turbulence: delocalised quasi strings and Majorana-like profiles}

In both the examples of Figs.~\ref{fig2} and \ref{fig3}, we considered a confined active nematic fluid. It is also of interest to study the behaviour of disclinations as quasiparticles in the bulk, without confinement. We focus on contractile activity in this case, to complement the extensile case studied under confinement. 
Such a contractile system could be experimentally realised with uniform actomyosin mixtures in 3D.

\begin{figure*}[t!]
\includegraphics[width=\textwidth]{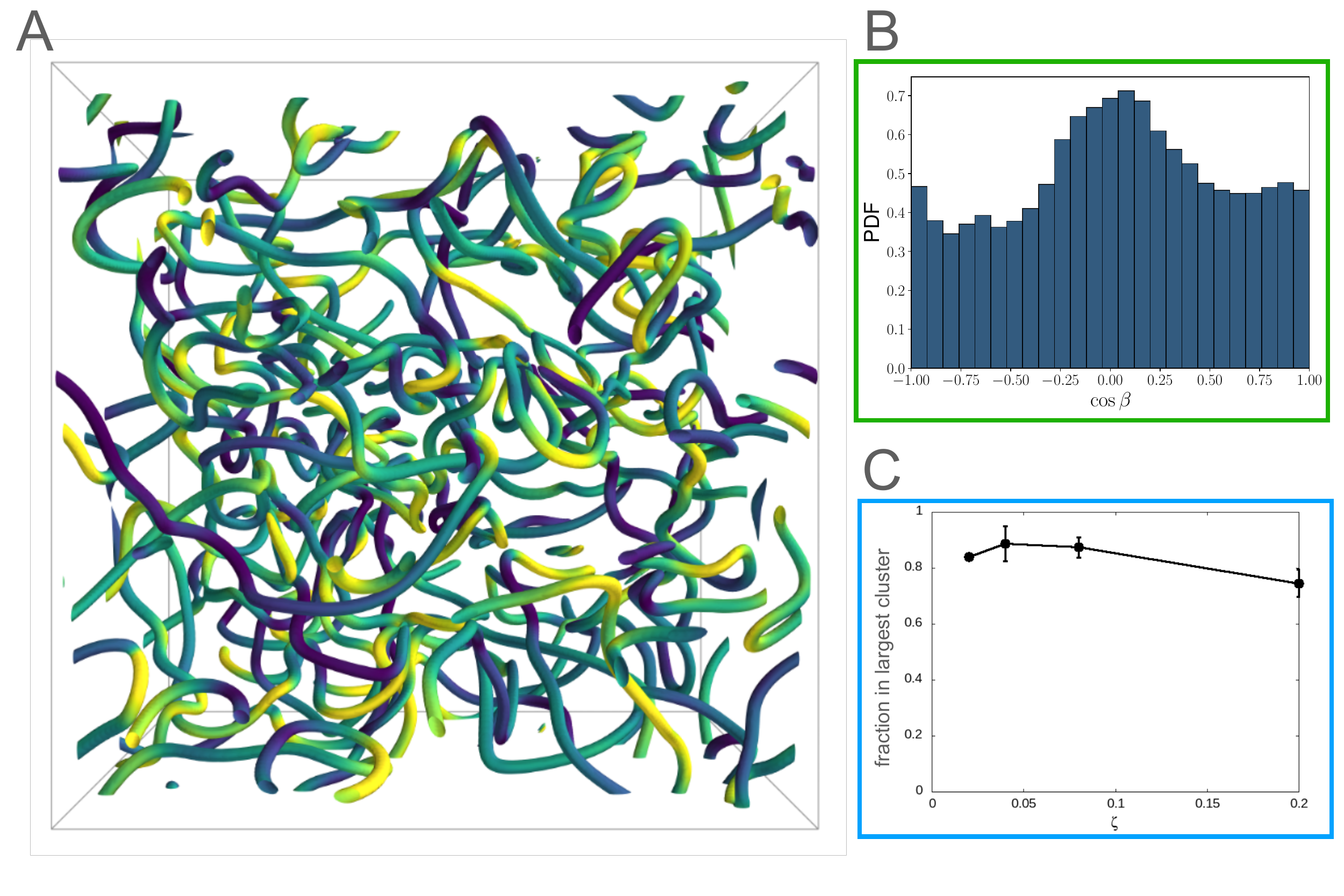}
\caption{(A) Snapshot of an active contractile nematic in bulk (parameter set given in Materials and Methods). 
Disclination lines are coloured according to the local $\cos{\beta}$ value. 
(B) Corresponding histogram of local values of $\cos{\beta}$. (C) A plot of the fraction of the disclination network in the largest connected defect cluster as a function of $\zeta$ 
(complete parameter set given in Material and Methods). Defects are identified here as regions in space with the order parameter smaller than $0.15$. 
} 
\label{fig4}
\end{figure*}

With periodic boundary conditions, the total charge of loops needs to be equal to $0$ modulo $2$. Similarly to the channel case, for low activity, the system is defect-free, whereas, after a sufficiently large activity, disclinations appear spontaneously in steady state. As a result, in the active turbulence phase, we observe a gel-like structure made of disclination loops and lines (Fig.~\ref{fig4}A). 

By analysing the topology and morphology of the disclination gels formed at different values of contractile activity $\zeta$, three main observations can be made. First, the local topological patterns (identified via the $\cos\beta$ values along disclination lines) are different to those of extensile nematics, where the local defect structure usually corresponds to a twist profile~\cite{duclos2020,binysh2020}. In the contractile case studied here, local profiles include more often triradii and comet-like defects (Fig.~\ref{fig4}B), corresponding to more pronounced Majorana-like local features. 
Second, 
virtually all the loops we have analysed are instead topologically trivial globally. 
This result is in line with observation on extensile nematics~\cite{duclos2020} and points to the key role played by boundaries in stabilising topologically charged loops. Third, we notice that an ensemble of short loops (corresponding to localised excitations, or quasiparticles) coexist with one or few very large percolating disclinations which wrap the system (corresponding to non-local excitations, Figs.~\ref{fig4}C and S2), similarly to what was observed in other recent simulations of active turbulence in 3D~\cite{digregorio2023}. 

We now translate these findings in terms of Majorana quasiparticles. At low activity, below the onset of active turbulence, Majorana-like loops are gapped and not observed. At sufficiently high activity, the gap closes, such that loop excitations become massless and are spontaneously formed and destroyed in steady state, leading to a statistical steady state. Here, most of the loops are topologically trivial and, hence are qualitatively akin to paired, and annihilating, Majorana quasiparticles. The few large wrapping disclinations~\cite{digregorio2023} correspond to extended quasiparticles, or quasistrings, and are reminiscent of the delocalised state in Fig.~\ref{fig3}B in confinement. As the fraction of disclination length in the largest (usually wrapping) disclination loop remains high for the whole range of activities which we have studied (Fig.~\ref{fig4}C), the system behaves as if the gap remains zero or very small throughout this range. Overall, these observations also suggest that active turbulence provides an intriguing topological phase where localised quasiparticles and delocalised topological excitations coexist dynamically for a large parameter range, similar to what happens in equilibrium close to a first-order transition.

\section*{Conclusions}

In summary, we have shown that a topologically charged disclination loop in a 3D active nematic liquid crystal provides a promising example of a classical topological quasiparticle whose nature is qualitatively similar to that of a Majorana particle. This is because, as the topological charge is conserved modulo $2$ in 3D nematics, a two-dimensional defect pattern can be smoothly transformed into its ``anti-defect'', i.e. the pattern which annihilates it: hence a loop with a local $-1/2$ profile is equivalent to a loop with a $+1/2$ profile. This is the defining feature of Majorana particles, which are equal to their antiparticles. More precisely, we can map a topologically trivial uncharged disclination loop to a pair of (annihilating) Majorana particles, and a topologically non-trivial charged loop to a Majorana quasiparticle in $1+1$D. These Majorana loops are gapped (or massive) excitations in passive unconfined nematic liquid crystals. Still, the gap can be closed -- equivalently the excitations become massless quasiparticles -- in the presence of confinement or activity. A single Majorana loop can be found in a double emulsion, where boundaries simultaneously provide a way to protect the loop topologically, and to stabilise a $-1/2$ triradius local defect profile.


We have shown that the analogy runs deeper, as we can find a way to qualitatively recreate a Kitaev chain in a confined quasi-1D active nematic fluid. Here, the activity provides a parameter similar to chemical potential in the Kitaev chain, as it can be tuned to trigger a transition between a trivial (passive) phase and a topological (active nematic) phase. In the former phase, at low activity,  Majorana excitations are gapped and hence absent. In the latter, at large activity, Majorana-like disclinations emerge spontaneously, as arcs absorbed onto the surface. Additionally, an extended helical disclination line arises for parameters for which the passive phase would be isotropic. The disclination handedness is selected by chiral symmetry breaking, and this state resembles non-local topological modes found in the Kitaev chain close to the transition between the trivial and topological phases.

Finally, extended topological excitations and local Majorana-like defect profiles also arise in unconfined active nematics, in the regime where active turbulence is observed. Here, we have found that systems with high contractile activity lead to the spontaneous formation of a melt of short uncharged loops, corresponding to paired Majorana quasiparticles, coexisting with one or few very extended disclinations, corresponding to non-local topological excitations, which can be interpreted as quasi strings. Such excitations provide the qualitative generalisation to an unconfined system of the helical disclination found in the Kitaev chain geometry. The fact that the vast majority of the loops we have analysed are uncharged suggests that Majorana-like loops require, or at least are much favoured by, the presence of boundaries, similar to the case of Majorana excitations in quantum condensed matter.

At this point, the analogy between active disclinations loops and Majorana fermions, whilst very suggestive, remains to a large degree qualitative, and it would be good to study in more detail the mathematical structure of the Clifford algebra and director (or vorticity) topological structures found in 3D active nematics. It would also be of interest to consider active cholesterics~\cite{whitfield2017,kralj2024}, as chirality further stabilises charged disclinations even in passive systems~\cite{fukuda2011}. We hope that our finding of a mapping between active nematics and Majorana physics will also stimulate further lab work, exploiting the larger structure of the topological Majorana-like excitations in these classical systems, which may facilitate their experimental investigation.

\section*{Materials and Methods}

\subsection*{Hydrodynamic equations of motion}

We first outline the general hydrodynamic equations we solved to obtain results for the three considered in the main text: (i) active emulsion (Fig.~\ref{fig2}); (ii) active nematic channel (Fig.~\ref{fig3}); (iii) active turbulence (Fig.~\ref{fig4}). System-specific details are given later, for each case individually.

We describe the physics of an active nematic fluid by considering two coarse-grained fields: (i) the fluid velocity ${\bf v}({\bf r}, t)$, and (ii) the tensorial order parameter ${\bf Q}({\bf r},t)$ (the ${\bf Q}$ tensor~\cite{degennes1993}) accounting for the ordering of a liquid crystal made of rod-like molecules.

The properties of the system in the passive limit are encoded by a free energy ${\mathcal F}$, whose density we call $f$, and which depends on the order parameter ${\bf Q}$. A form of the free energy density which is appropriate to describe a nematic liquid crystal is the following (note summation over repeated indices is implied):
\begin{eqnarray}\label{free}
&f_{\rm LC}&=
\frac{A_0}{2}\left(1-\frac{\gamma}{3}\right)Q_{\alpha\beta}^2-\frac{A_0\gamma}{3}Q_{\alpha\beta}Q_{\beta\gamma}Q_{\gamma\alpha}\nonumber\\&&+\frac{A_0\gamma}{4}(Q^2_{\alpha\beta})^2+\frac{K}{2}(\partial_{\gamma}Q_{\alpha\beta})^2, 
\end{eqnarray}
where Greek indices denote the Cartesian components of tensors.
The first three terms on the right-hand side of Eq.~(\ref{free}) constitute the bulk free energy density and are the first terms in an expansion over powers of the ${\bf Q}$ tensor, where $A_0$ is a positive constant and $\gamma$ a parameter controlling the transition between isotropic and nematic phase (at $\gamma<\gamma_c=2.7$ and $\gamma>\gamma_c$ respectively). 
The energetic cost due to liquid crystal distortions is controlled by the gradients of $Q_{\alpha\beta}$ and is proportional to $K$, the elastic constant.

The time evolution of the ${\bf Q}$ tensor is governed by the Beris-Edwards equation~\cite{beris1994}
\begin{equation}\label{beris_eqn}
(\partial_t + {\bf v}\cdot\nabla){\bf Q}-{\bf S}({\bf W},{\bf Q})=\Gamma{\bf H},
\end{equation}
where $\Gamma$ is a collective rotational diffusion constant and ${\bf H}$ is the molecular field, which is given by
\begin{equation}
{\bf H}=-\frac{\delta {\cal F}}{\delta {\bf Q}}+({\bf I}/3)Tr\frac{\delta {\cal F}}{\delta {\bf Q}},
\end{equation}
with ${\bf I}$ the identity matrix. 
The first term on the left-hand side of Eq.~(\ref{beris_eqn}) represents the material derivative accounting for the time dependence of a quantity advected by the fluid velocity ${\bf v}$. The second term accounts for the rotation and stretching of the rod-like molecules of the liquid crystals due to flow gradients. This is given by~\cite{beris1994}
\begin{eqnarray}
{\bf S}({\bf W},{\bf Q})&=&(\xi{\bf D}+{\bm\omega})\cdot({\bf Q}+{\bf I}/3) \nonumber\\&+&
({\bf Q}+{\bf I}/3)\cdot(\xi{\bf D}-{\bm\omega}) \nonumber\\
&-&2\xi({\bf Q}+{\bf I}/3)Tr({\bf Q}\cdot{\bf W}).
\end{eqnarray}
Here $Tr$ denotes the tensorial trace, while ${\bf D}=({\bf W}+{\bf W}^T)/2$ and ${\bm\omega}=({\bf W}-{\bf W}^T)/2$ are the symmetric and anti-symmetric part of the velocity gradient tensor $W_{\alpha\beta}=\partial_{\beta}v_{\alpha}$. The parameter $\xi$ determines whether the liquid crystal is flow-aligning or flow-tumbling. 

The fluid velocity ${\bf v}$ evolves according to the Navier-Stokes equation,
\begin{equation}\label{nav_stok_eqn}
\rho(\partial_t{\bf v}+{\bf v}\cdot \nabla{\bf v})=-\nabla p + \nabla\cdot ({\bm \sigma}^{visc}+{\bm \sigma}^{lc}
+{\bm \sigma}^{act}),
\end{equation}
complemented by the incompressibility condition $\nabla\cdot{\bf v}=0$. In Eq.~(\ref{nav_stok_eqn}), $\rho$ is the fluid density and $p$ is the hydrodynamic pressure. The stress ${\bm \sigma}$ is broken up into three terms. The viscous contribution, ${\bm \sigma}^{visc}$, is given by
\begin{equation}
\sigma_{\alpha\beta}^{visc}=\eta^{visc}(\partial_{\alpha}v_{\beta}+\partial_{\beta}v_{\alpha}),
\end{equation}
where $\eta^{visc}$ is the shear viscosity of the fluid. 
The elastic stress due to liquid crystal deformations,  ${\bm \sigma}^{lc}$, reads
\begin{eqnarray}
\sigma_{\alpha\beta}^{lc}&=&-\xi H_{\alpha\gamma}(Q_{\gamma\beta}+\frac{1}{3}\delta_{\gamma\beta})-\xi(Q_{\alpha\gamma}+\frac{1}{3}\delta_{\alpha\gamma})H_{\gamma\beta}\nonumber\\
&+&2\xi(Q_{\alpha\beta}+\frac{1}{3}\delta_{\alpha\beta})Q_{\gamma\mu}H_{\gamma\mu}+Q_{\alpha\gamma}H_{\gamma\beta}-H_{\alpha\gamma}Q_{\gamma\beta} \nonumber\\
&-&\partial_{\alpha}Q_{\gamma\mu}\frac{\partial f}{\partial(\partial_{\beta}Q_{\gamma\mu})}.
\end{eqnarray}
The third term, ${\bm \sigma}^{act}$, is the active stress given by \cite{marchetti2013}
\begin{equation}
\sigma_{\alpha\beta}^{act}=-\zeta Q_{\alpha\beta},
\end{equation}
where $\zeta$ is the activity parameter, positive for extensile materials and negative for contractile ones. 

There are two key dimensionless parameters in all systems we considered. One is the reduced activity $\tilde{A}$, which is the ratio between a typical lengthscale $l$ (droplet radius or channel width or system size, according to the system of interest) and the active lengthscale $l_a=\sqrt{K/\zeta}$~\cite{marchetti2013}. The other is the reduced temperature $\tau=\frac{27(1-\gamma/3)}{\gamma}$, which determines whether the system in the passive phase is isotropic ($\tau>1$) or nematic ($\tau<1$). A third dimensionless parameter determines the strength of anchoring at boundaries in systems with confinement.

\subsection*{Active emulsion}

To describe an active emulsion, in addition to the velocity field and the ${\bf Q}$ tensor, a set of passive scalar phase fields  $\phi_i({\bf r}, t), i = 1,2,3$ is required, to capture the density of each droplet. The active droplet in Fig.~\ref{fig2} is associated with the phase field $\phi_1$, whereas the two passive droplets are with $\phi_{2,3}$.

The free energy density in Eq.~(\ref{free}) also needs to be complemented by an additional term to describe the behaviour of the phase fields; this reads, 
\begin{eqnarray}\label{freephi}
&f_{\phi}&=\frac{a}{4}\sum_{i=1}^3 \left[\phi_i^2(\phi_i-\phi_0)^2\right]\nonumber+\frac{k}{2}\sum_{i=1}^3(\nabla\phi_i)^2\\ \nonumber \\&&+\epsilon_{\phi}\sum_{i\ne j; i,j=1}^3\phi_i^2\phi_j^2 
+W_{\phi}\sum_{i=1}^3\left[\partial_{\alpha}\phi_i Q_{\alpha\beta}\partial_{\beta}\phi_i\right].
\end{eqnarray}
The first term in Eq.~(\ref{freephi}), multiplied by the positive constant $a$, represents a double well potential which ensures the existence of two coexisting minima at \textcolor{black}{$\phi_{1,2,3}=\phi_0=2$ (this represents the inside of droplets $2$  and $3$ and the outside of droplet $1$) and $\phi_{1,2,3}=0$ (inside of droplet $1$ and outside the droplets $2$ and $3$)}. The second term, multiplied by the elastic constant $k$, controls the interfacial energy. 
The constants $a$ and $k$ together determine
the surface tension $\sigma=\sqrt{{8ak/9}}$ and the interface thickness $\xi_{\phi}=\sqrt{2k/a}$ of the droplets.
The third contribution is a soft-core repulsion, whose magnitude is controlled by the positive constant $\epsilon_{\phi}$. 
In $f_{\rm LC}$ [Eq.~(\ref{free})], the coefficient $\gamma$ is now a function of the phase fields. Following previous works \cite{carenza2019}, we set $\gamma=\gamma_0+\gamma_s\sum_{i=1}^3\phi_{i}$, where  $\gamma_0$ and $\gamma_s$ control the boundary of the coexistence
region. \textcolor{black}{Note that $\gamma$ depends on the sum of $\phi_i$, since the liquid crystal is confined solely within the layer where $\sum_{i=1}^3\phi_i=0$}. The anchoring of the director at the droplet interface is described by the last term, where $W_{\phi}$ is the anchoring strength ($W_{\phi}<0$ corresponds to normal, or homeotropic, anchoring at droplet surfaces).

The dynamics of the scalar fields $\phi_i$ ($i=1,2,3$) obeys a Cahn-Hilliard equation
\begin{equation}\label{cahn_eqn}
\partial_t \phi_i+{\bf v}\cdot{\nabla\phi_i}=M\nabla^2\mu_i,
\end{equation}
where $M$ is the mobility and $\mu_i=\frac{\delta{\cal F}}{\delta\phi_i}$ is the chemical potential.

A further contribution to the stress stems from interfacial stress 
and is given by
\begin{equation}
\sigma^{int}_{\alpha\beta}=\sum_{i=1}^{3}\left[\left(f-\phi_i\frac{\delta{\cal F}}{\delta\phi_i}\right)\delta_{\alpha\beta}-\frac{\partial f}{\partial(\partial_{\beta}\phi_i)}\partial_{\alpha}\phi_i\right].
\end{equation}

As in previous works~\cite{carenza2019}, we used a 3D hybrid lattice Boltzmann (LB) method which solves Eq.~(\ref{cahn_eqn}) and Eq.~(\ref{beris_eqn}) via a finite difference scheme while 
Eq.~(\ref{nav_stok_eqn}) was solved by a lattice Boltzmann approach. Parameters used to generate the configuration shown in Fig.~\ref{fig2} are: $\sigma\simeq 0.035$ $\xi_{\phi}\simeq 5.29$, $M=0.1$, $R_1/l_a\simeq 17.75$ (where $R_1=32$ is the initial radius of the active droplet, whereas, $R_2=R_3=8$ are the initial radii of the isotropic cores), dimensionless anchoring strength $W_{\phi}R/(K \xi_{\phi})\simeq -4.65$, $A_0=0.12$, $\gamma_0=2.85$, $\gamma_s=-0.25$ (so that in the active nematic region $\tau\simeq 0.47$).

\subsection*{Active channel}

Hybrid LB active channel simulations (Fig.~\ref{fig3}) were confined by impermeable no-slip planar walls. The free energy density in Eq.~(\ref{free}) is complemented by an anchoring free energy term with minima at $\qtens=S_0\left({\bf \nu}\otimes{\bf \nu}-\frac{1}{3}\identitytens\right)$, with $S_0$ the preferred degree of surface order and ${\bf{\nu}}$ an arbitrary orientation parallel to the surface. This anchoring free energy reads~\cite{fournier2005} 
\begin{eqnarray}
    \label{eq:planardegenerate}
    {\mathcal F}_\mathrm{anch} & = &  \int dS \large[ W_1\left(\tilde{\qtens}-\tilde{\qtens}^{\perp}\right):\left(\tilde{\qtens}-\tilde{\qtens}^\perp\right)  \nonumber \\ 
    &+& W_2\left(\tilde{\qtens}:\tilde{\qtens}-S_{0}^2\right) \large].
\end{eqnarray}
The first term imposes {\it alignment} on the tangent plane with an energy cost $W_1$. Here, $\tilde{\qtens}=\qtens+\frac{1}{3}S_0\identitytens$ and $\tilde{\qtens}^\perp=\mathbf{P}\cdot\tilde{\qtens}\cdot\mathbf{P}$ in terms of the projection operator $\mathbf{P}$ on the plane perpendicular to the surface normal.
The second term sets the {\it degree} of order ($S_0$) on the plane, with energy cost $W_2$. In simulations analysed here, $W_1=W_2$

The active nematic fluid is confined within a parallelepiped of size $L_x \times L_y\times L_z$. For both the states in Fig.~\ref{fig3}, $\rho = 1$, $\eta^{visc}=4/3$, $\Gamma = 0.3375$, $\xi = 1$, $L_y=L_z=25$. Additionally, for the vortex lattice (Fig.~\ref{fig3}A), $L_x = 130$, 
$A_0=1$, $\tau=0$ ($\gamma=3$), $\sqrt{\zeta/K}L_{y,z}=22$, $W_{1,2}L_{y,z}/K=250$. For the double helical state (Fig.~\ref{fig3}B), $L_x=200$, $A_0=0.1$, $\tau\simeq 1.120$ ($\gamma=2.668$), $\sqrt{|\zeta|/K}L_{y,z}=25$, and dimensionless anchoring strength at the walls $W_{1,2}L_{y,z}/K=1.5$. 



\subsection*{Active turbulence} 

Active turbulence simulations (Fig.~\ref{fig4}) are based on Eq.~(\ref{free}), and were performed in a cubic box of size $L=256 \sqrt{K/A_0}$ with periodic boundary conditions. 
Simulations were solved numerically with an in-house MPI-parallel fully-dealiased pseudo-spectral code developed within the Dedalus framework~\cite{Burns2020}. All fields are represented by a triple-Fourier spectral decomposition with the spectral resolution being given by the number of modes in each direction $(N_x,N_y,N_z)$. Production runs are carried out using $N_x=N_y=N_z=256$ comprising approximately $1.5\cdot 10^8$ degrees of freedom. Temporal discretisation employs the semi-implicit backward differentiation scheme of order four \cite{Wang2008} with a time step of $0.01$. All simulations were evolved for at least $10^4$ time units in a statistically steady state, where time is measured in units of $(\Gamma A_0)^{-1}$. Parameters corresponding to Fig.~\ref{fig4} are $\Gamma=0.3$, 
$\xi=0.7$, $A_0=1.5$, $\tau=0$ ($\gamma=3$), $K=0.06$, $\zeta=-0.08$ (Figs.~\ref{fig4}A,B) and $\zeta \in \{-0.02,-0.2\}$ (Fig.~\ref{fig4}C). Therefore the ratio between the system size and the active lengthscale is $\simeq 59.12$
in Figs.~\ref{fig4}A,B, and between $29.56$ and $93.48$
in Fig.~\ref{fig4}C. 

\subsection*{Defect analysis}

Within the theory of Majorana particles and bound states, the Majorana polarisation and Pfaffian charge density~\cite{sticlet2012} provide useful quantities to identify topologically non-trivial states associated with Majorana physics. For our nematic counterpart, 
a qualitatively analogous quantity is the disclination line tensor~\cite{schimming2022}, which gives a way to visualise the local defect profile of a liquid crystalline pattern in a simple way~\cite{head2024}, determining if the profile on a disclination segment is triradius-like, comet-like, or twist-like (Fig.~\ref{fig1}). 
The tensor is constructed from derivatives of the Q-tensor,
\begin{align}
    D_{ij}=\epsilon_{i\mu\nu}\epsilon_{jlk}\partial_l Q_{\mu\alpha}\partial_k Q_{\nu\alpha}
\end{align}
where $i,j,k,\alpha,\mu,\nu$ are tensor indices with applied summation convention. This form is useful due to the interpretation as the dyad composing of the local line tangent $\mathbf{T}$ and the rotation vector $\mathbf{\Omega}$
\begin{equation}
    \label{Dij}
    D_{ij} = s(\mathbf{r}){\Omega}_{i}{T}_j.
\end{equation}
where $s(\mathbf{r})$ is a positive scalar field that is maximum at the disclination core. Defects are identified as isosurfaces of $s(\mathbf{r})=s_\mathrm{cut}$. For the double emulsion $s_\mathrm{cut}=0.033$, double helix $s_\mathrm{cut}=0.023$, vortex lattice $s_\mathrm{cut}=0.1$ and active turbulence $s_\mathrm{cut}=0.01$. For Fig.~\ref{fig2}B (topologically charged Majorana-like loop) and Fig.~\ref{fig3}Bi (double helix configuration), topological patterns are visualised as a tube that connects an ordered sequence of points along the defect loop. 

Extracting $\mathbf{\Omega}$ and $\mathbf{T}$ use the methods outlined in \cite{schimming2022}, ensuring that the vectors are continuous along the loop and have the correct relative sign, set by sgn$(\mathbf{\Omega}\cdot\mathbf{T})=$sgn(Tr($D_{ij}$)).
Visualisations of disclinations and disclination networks use the Mayavi library~\cite{ramachandran2011}.


\subsection*{Acknowledgements}

{For the purpose of open access, the author has applied a Creative Commons Attribution (CC BY) licence to any Author Accepted Manuscript version arising from this submission.}




\setcounter{figure}{0}    
\newcommand{\SIFig}[1]{Fig. S#1}
\renewcommand{\thefigure}{S\arabic{figure}}

\newpage


\begin{figure*}[h!]
\centering
\includegraphics[width=\textwidth]{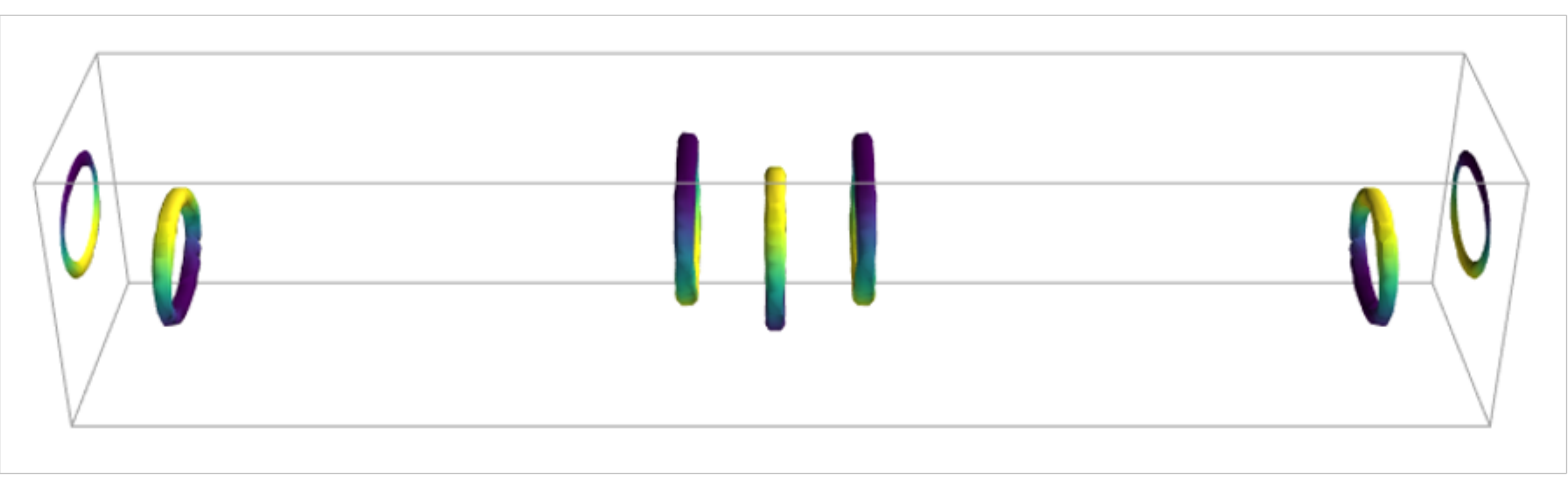}
\caption{Patterns of disclination loops (coloured according to the local $\cos{\beta}$ value, as in Fig.~1 of the main text), for the vortex lattice (whose steady state configuration is shown in Fig. 3Ai).}
\end{figure*}

\begin{figure*}[h!]
\centering
\includegraphics[width=\textwidth]{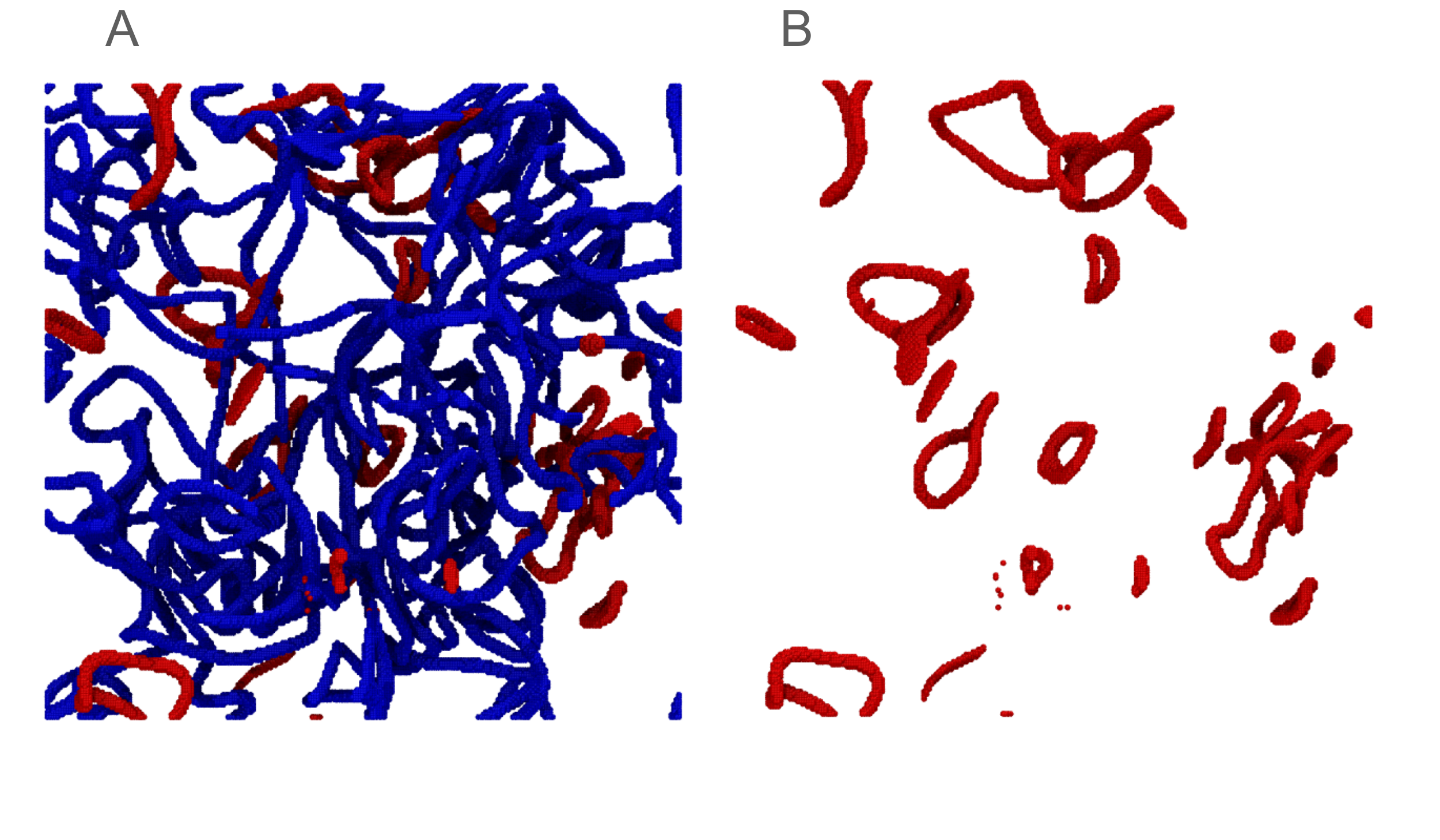}
\caption{A. Disclination line patterns coloured according to whether a defect (order parameter smaller than $0.18$)  is in the largest cluster (blue) or not (red). Parameters correspond to those of Fig. 3C, for $\zeta=-0.08$. B. Same as A but without the largest connected component.}
\end{figure*}


\begin{thebibliography}{99}
\bibitem{mermin1979} N.~D. Mermin, {\it Rev. Mod. Phys.} {\bf 51}, 591-648 (1979).
\bibitem{mermin1989} N.~D. Mermin, {\it Rev. Mod. Phys.} {\bf 61}, 385 (1989).
\bibitem{chuang1991} I. Chuang, R. Durrer, N. Turok, and B. Yurke, {\it Science} {\bf 251}, 1336-1342 (1991).
\bibitem{lavrentovich2001} O.~D. Lavrentovich and M. Kleman, ``Cholesteric Liquid Crystals: Defects and Topology''in H.-S. Kitzerow, C. Bahr (Eds.), {\it Chirality in Liquid Crystals}, Springer, New York (2001).  
\bibitem{ackerman2017} P.~J. Ackerman and I.~I. Smalyukh, {\it Phys. Rev. X} {\bf 7}, 011006 (2017).
\bibitem{poy2022} G. Poy, A.~J. Hess, A.~J. Seracuse, S. Zumer, and I.~I. Smalyukh, {\it Nat. Photonics} {\bf 16}, 454-461 (2022).
\bibitem{wu2022} J.-S. Wu and I.~I. Smalyukh, {\it Liq. Cryst. Rev.}, DOI: 10.1080/21680396.2022.2040058 (2022).
\bibitem{carenza2022} L.~N. Carenza, G. Gonnella, D. Marenduzzo, G. Negro, and E. Orlandini, {\it Phys. Rev. Lett.} {\bf 128}, 027801 (2022).
\bibitem{shankar2022}
S. Shankar, A. Souslov, M.~J. Bowick, M.~C. Marchetti, and V. Vitelli, {\it Nat. Rev. Phys.} {\bf 4}, 380 (2022).
\bibitem{bowick2022}
M.~J. Bowick, N. Fakhri, M.~C. Marchetti, and S. Ramaswamy,  {\it Phys. Rev. X} {\bf 12}, 010501 (2022).
\bibitem{kitaev2001} A.~Y. Kitaev, {\it Physics-Uspekhi} {\bf 44}, 131 (2001). 
\bibitem{qi2011} X.~L. Qi and S.~C. Zhang, {\it Rev. Mod. Phys.} {\bf 83}, 1057 (2011).
\bibitem{alicea2012} J. Alicea, {\it Rep. Prog. Phys.} {\bf 75}, 076501 (2012).
\bibitem{wolfle2018} P. W{\"o}lfle, {\it Rep. Prog. Phys.} {\bf 81}, 032501 (2018).
\bibitem{majorana1937} E. Majorana, {\it Nuovo Cimento} {\bf 5}, 171 (1937).
\bibitem{copar2013} S. Copar and S. Zumer, {\it Proc. R. Soc. A} {\bf 469}, 20130204 (2013). 
\bibitem{mohanta2021} N. Mohanta, S. Okamoto, and E. Dagotto, {\it Commun. Physics} {\bf 4}, 163 (2021).
\bibitem{ramaswamy2010} S. Ramaswamy, {\it Annu. Rev. Condens. Matter Phys.} {\bf 1} 323 (2010).
\bibitem{marchetti2013} M.~C. Marchetti, J.~F. Joanny, S. Ramaswamy, T.~B. Liverpool, J. Prost, M. Rao, and R.~A. Simha, {\it Rev. Mod. Phys.} {\bf 85}, 1143 (2013).
\bibitem{doostmohammadi2018}
A. Doostmohammadi, J. Ign{\'e}s-Mullol, J.~M. Yeomans, and F. Sagu{\'e}s, {\it Nat. Commun.} {\bf 9}, 3246 (2018).
\bibitem{duclos2020} G. Duclos, R. Adkins, D. Banerjee, M.~S.~E. Peterson, M. Varghese, I. Kolvin, A. Baskaran, R.~A. Pelcovits, T.~R. Powers, A. Baskaran, F. Toschi, M.~F. Hagan, S. J. Streichan, V. Vitelli, D.~A. Beller, and Z. Dogic,
{\it Science} {\bf 367}, 1120 (2020).
\bibitem{binysh2020} J. Binysh, Z. Kos, S. Copar, M. Ravnik, and G.~P. Alexander, {\it Phys. Rev. Lett.} {\bf 124}, 088001 (2020).
\bibitem{kauffman2022} L.~H. Kauffman,arXiv:2202.13592 (2022).
\bibitem{kitaev2003} A. Kitaev, {\it Ann. Phys.} {\bf 303}, 2 (2003). 
\bibitem{lian2018} B. Lian, X.-Q. Sun, A. Vaezi, X.-L. Qi, and S.-C. Zhang, {\it Proc. Natl. Acad. Sci. USA} {\bf 115}, 10938 (2018).
\bibitem{kos2022} Z. Kos and J. Dunkel, {\it Sci. Adv.} {\bf 8}, eabp8371 (2022).
\bibitem{alert2022} R. Alert, J. Casademunt, and J.~F. Joanny, {\it Annu. Rev. Condens. Matter Phys.} {\bf 13}, 143 (2022).
\bibitem{giomi2015} L. Giomi, {\it Phys. Rev. X} {\bf 5}, 031003 (2015).
\bibitem{alert2020} R. Alert, J.~F. Joanny, and J. Casademunt, {\it Nat. Phys.} {\bf 16}, 682-688 (2020).
\bibitem{pal2011} P.~B. Pal, {\it Am. J. Phys.} {\bf 79}, 485-498 (2011).
\bibitem{schimming2022} C.~D. Schimming and J. Vinals, {\it Soft Matter} {\bf 18}, 2234-2244 (2022).
\bibitem{head2024} L.~C. Head, C. Dor{\'e}, R.~R. Keogh, L. Bonn, Lasse, G. Negro, D. Marenduzzo, A. Doostmohammadi, K. Thijssen, T. L{\'o}pez-Le{\'o}n, and T.~N. Shendruk, {\it Nat. Phys.} (2024), https://doi.org/10.1038/s41567-023-02336-5.
\bibitem{adriano2024} G. Negro, L.~C. Head, L.~N. Carenza, T.~N. Shendruk, D. Marenduzzo, G. Gonnella, and A. Tiribocchi, arXiv:2402.02960.
\bibitem{stark2001} H. Stark, {\it Phys. Rep.} {\bf 351}, 387 (2001).
\bibitem{keogh2022} R.~R. Keogh, S. Chandragiri, B. Loewe, T. Ala-Nissina, S.~P. Thampi, and T.~N. Shendruk, {\it Phys. Rev. E} {\bf 106}, L012602 (2022).
\bibitem{keber2014} F.~C. Keber, E. Loiseau, T. Sanchez, S.~J. DeCamp, L. Giomi, M.~J. Bowick, M.~C. Marchetti, Z. Dogic, and A.~R. Bausch, {\it Science} {\bf 345}, 1135-1139 (2014).
\bibitem{digregorio2023} P. Di Gregorio, C. Rorai, I. Pagonabarraga, and F. Toschi, arXiv:2307.10103.
\bibitem{whitfield2017} C.~A. Whitfield, T.~C. Adhyapak, A. Tiribocchi, G.~P. Alexander, D. Marenduzzo, and S. Ramaswamy, {\it Eur. Phys. J. E} {\bf 40}, 50 (2017).
\bibitem{kralj2024} N. Kralj, M. Ravnik, and Z. Kos, arXiv:2402.10020.
\bibitem{fukuda2011} J.-I. Fukuda and S. Zumer, {\it Phys. Rev. Lett.} {\bf 106}, 097801 (2011).
\bibitem{degennes1993} P.~G. de Gennes and J. Prost, {\it The physics of liquid crystals}, Oxford University Press, Oxford (1993).
\bibitem{beris1994} A.~N. Beris and B.~J. Edwards, {\it Thermodynamics of flowing systems with internal microstructure}, Oxford University Press, USA (1994).
\bibitem{carenza2019} L.~N. Carenza, G. Gonnella, D. Marenduzzo, and G. Negro, {\it Proc. Natl. Acad. Sci. USA} {\bf 116}, 22065 (2019).
\bibitem{fournier2005} J.-B. Fournier and P. Galatola, {\it EPL} {\bf 72}, 403 (2005).
\bibitem{Burns2020} K.~J. Burns, G.~M. Vasil, J.~S. Oishi, D. Lecoanet, and B.~P. Brown, 
{\it Phys. Rev. Res.} {\bf 2}, 023068 (2020).
\bibitem{Wang2008}
D. Wang and S.~J. Ruuth, 
{\it J. Comput. Math.} {\bf 26}, 838 
  (2008).
\bibitem{sticlet2012} D. Sticlet, C. Bena, and P. Simon, {\it Phys. Rev. Lett.} {\bf 108}, 096802 (2012).
\bibitem{ramachandran2011} P. Ramachandran and G. Varoquaux, {\it Comput. Sci. Eng.} {\bf 13}, 40 (2011).
\end{thebibliography}
\end{document}